\documentclass[rmp,showkeys,nofootinbib,twocolumn]{revtex4}
\usepackage{graphicx}
\ifnum\lefthyphenmin<2\lefthyphenmin=2\fi
\ifnum\righthyphenmin<2\righthyphenmin=2\fi
\bibpunct[, ]{(}{)}{;}{a}{,}{,}
\newcommand{\vS}{\vec{S}}
\newcommand{\talpha}{\widetilde{\alpha}}

\newcommand{\cA}{{\cal A}}
\newcommand{\br}{\langle r\rangle}
\newcommand{\eps}{\varepsilon}
\newcommand{\bph}{\overline{\phi}}
\begin{document}
\title{On the Selection and Evolution of Regulatory DNA Motifs}
\author{Ulrich Gerland\footnote{email: gerland@physics.ucsd.edu}}
\author{Terence Hwa\footnote{email: thwa@ucsd.edu}}
\affiliation{Department of Physics, University of California at San Diego
La Jolla, CA 92093-0319}
\affiliation{Institute for Theoretical Physics, University of California, 
Santa Barbara, CA 93106-4030}
\date{\today}
%
%
\begin{abstract}
The mutation and selection of regulatory DNA sequences is presented 
as an ideal model system of molecular evolution where genotype, phenotype,
and fitness can be explicitly and independently characterized.
In this theoretical study, we construct an explicit model 
for the evolution of regulatory sequences, making use of the known
biophysics of the binding of regulatory proteins to DNA sequences,
under the assumption that fitness of a sequence depends only on
its binding affinity to the regulatory protein.
The model is confined to the mean field (i.e., infinite population
size) limit.
Using realistic values for all parameters, we determine the minimum fitness
advantage needed to maintain a binding sequence, demonstrating explicitly
the ``error threshold'' below which a binding sequence cannot survive the 
accumulated effect of mutation over long time. The commonly 
observed ``fuzziness'' in binding motifs arises naturally as a consequence
of the balance between selection and mutation in our model.
In addition, we devise a simple model for the evolution of multiple
binding sequences in a given regulatory region. We find the number of
evolutionarily stable binding sequences to increase in a step-like fashion 
with increasing fitness advantage, if multiple regulatory proteins
can synergistically enhance gene transcription.
We discuss possible experimental approaches to resolve open questions 
raised by our study. 
\end{abstract}
\keywords{transcription regulation, mutation-selection model, 
          error threshold, regulatory sequences}
\maketitle
\section{Introduction}
The regulation of gene expression involves many different 
proteins known as transcription factors which bind passively 
to specific sites on the genomic DNA (see, e.g. \citet{gerhart}).
In bacteria, each such site (called `operator') typically 
consists of a contiguous sequence of $20-30$ nucleotides
which binds a specific transcription factor with much higher affinity
than would a random DNA sequence of comparable length 
\citep{vonhippel79,stormo}. 
Known examples of different operators for the same factor usually 
differ from the maximum affinity binding sequence in a number of  
positions, typically in as many as $20\%$ to $30\%$ of
the significant positions that contribute most to the 
specificity of the interaction. 
The ensemble of viable binding sequences is collectively referred 
to as the binding ``motif'' for a factor; its ``fuzziness'' creates 
a difficult computational problem for the prediction of binding 
sites via informatic methods (see, e.g., \citet{stormo89,gibbs};  
and references therein). 
In many known cases, a {\em single} regulatory region contains 
{\em multiple} operators for the same factor, each of which deviates 
from the maximum affinity binding sequence.  

Why are the motifs fuzzy? One possible scenario is that the binding affinity
of each operator is tuned evolutionarily to maximize the function
of each regulated gene or operon.  An alternative scenario is that 
the function is insensitive to the detail of the binding affinity
as long as it is above some threshold. In the former case, fuzziness
in the binding arises due to the particular distribution of functional
requirment. In the latter case, binding sequences in different regulatory
regions are deemed ``equal'', and fuzziness results from maximizing the
sequence ``entropy''. While anecdotal examples of both
cases are known, understanding whether either case dominates in biology
is of interest not only for its own sake, but also very important for
the choice of proper informatics tools for motif finding.
 In this paper, we describe a detailed
theoretical study of the latter case from an evolutionary perspective, 
recognizing that as with any other portion of the genome, 
the binding sequences are subject to the opposing forces of mutation 
and selection over evolutionary timescales.  
In particular, we address the quantitative question of how large a 
selective advantage the presence of a binding motif needs to provide, 
in order to guarantee its survival against mutations, and how large
an advantage before multiple motifs are justified.
To make the study concrete and explicit, we will mostly confine 
the discussion to gene regulation in bacteria or phages, and focus on 
the binding of one specific transcription factor to
its operator(s) in the regulatory region of one specific gene or operon.
We will not treat the interactions among different factors, since
in bacteria such as {\it E.~coli}, the majority of genes are regulated
by a single factor~\citep{gralla}.

Another motivation for our study is that
the evolution of transcription factor binding motifs seems to be a 
well-suited starting point for an attempt to establish a link between the 
microscopic molecular mechanisms in the cell and the ``macroscopic'' 
principles of evolution: In general, the most important ingredient 
in an evolutionary study is to relate the genotype on which mutation 
acts to the fitness of the organism on which selection acts through 
some quantifiable phenotype. 
This relation is particularly simple for the operator binding problem 
at hand, where a natural choice of the phenotype is the binding
probability of the transcription factor to the operator.
Regardless of whether the factor acts as an activator by attracting 
a polymerase to transcribe the gene, or as a repressor to block 
transcription, it can function only when it is bound to its operator.
The fraction of time an operator is occupied in equilibrium
is given by the binding probability. 
To regulate the transcription of the gene, e.g., in reaction to a change 
in the environmental conditions or in order to trigger a different phase 
of the cell cycle, the cell changes the factor-operator binding 
probability by varying the concentration of the (activated) factor inside the cell. 
The concentration may vary from practically zero in the ``OFF-state'' to 
typically several hundred copies per cell in the ``ON-state''.
We will make the reasonable (but critical) assumption that
the fitness gain an operator contributes to the organism 
depends solely on the binding probability $P$ in the ON-state, 
with the value of $P$ itself determined by the actual sequence of the 
operator through the binding energetics.

For a few exemplary transcription factors, the variation in binding 
affinity upon mutation of the binding sequence has been studied in great 
detail experimentally~\citep{fields,takeda,sarai,oda}. 
In particular, Fields, Stormo, and coworkers have shown for the case 
of the mnt repressor that its binding (free) energy is approximately a 
sum over {\em independent} contributions from each of the nucleotide 
positions in the binding sequence~\citep{fields}. 
Typically, only $10\sim 15$ positions in a binding site have a strong 
preference for specific nucleotides, while the other positions do not 
contribute significantly to the binding energy. 
Known binding sequences display a fuzziness of up to $3\sim 4$ mismatches 
in these significant positions. 
A useful simplified `two-state model' for transcription factor binding 
is obtained by taking only the significant bases into account and 
assigning to each of them the same binding energy $\eps$, i.e. a 
match (to the optimal binding sequence) 
is favored by an energy difference $\eps$ over a mismatch.
This model, introduced long ago by \citet{vonhippel86}, 
takes into account the effect of 
sequence-specific binding by a single parameter $\eps$, 
without reference to detailed binding energies which 
have not yet been measured for most transcription factors.

Based on the two-state model and our assumption on the contribution
of the  binding of the factor towards fitness, we construct an 
explicit theory for the evolution of the binding sequences. 
Within the mean-field approach originally proposed by Eigen in the 
context of quasispecies evolution~\citep{eigen71,eigen89}, we 
characterize the balance between the opposing forces of selection 
and mutation quantitatively.
We determine the critical selection pressure needed to keep 
a motif from mutating away, and show how the fuzziness in the motifs 
arises naturally above the selection threshold. 
We further apply the theory to investigate the frequently observed occurrence 
of multiple motifs in a given regulatory region, and elaborate on 
various plausible causes. Towards the end, we will provide 
extended discussions on experimental approaches to pursue the open questions
suggested by this study.
\section{Model and Equations}
We focus on the operator sequence located in the regulatory region of 
a gene of interest. 
By assumption, this gene is regulated by a single transcription factor.
Let $\vS = \{S_1, S_2, ..., S_L\}$ denote 
the $L$ significant nucleotides of the operator which specify 
transcription factor binding.
We will keep the alphabet size, $\cA$, as a variable in our equations, 
since, as we will see below, this facilitates the intuitive 
understanding of certain dependencies; however $\cA=4$ and 
$S_i \in \{{\tt A,C,G,T}\}$ is the only case of interest here. 
To describe the evolution of $\vS$ in a population of bacteria or phages, we  
need to specify the action of selection and mutation. 

\vspace{24pt}
\noindent{\bf Selection mechanism}
\vspace{12pt}

It should be clear that gene regulation is only needed in the
presence of a {\em changing} cellular state, triggered either internally,
e.g. cell cycle, or externally through a change in the environment.
Hence to study the fitness of a regulatory mechanism, we must invoke at 
least two different states. 
Selection arises when the growth rate of an organism 
depends on the probability $P_{\vS}$ that the factor binds
to the sequence $\vS$ in the state that prefers factor binding 
(the ``ON-state''). 
For the sake of concreteness, let us consider an environment that 
oscillates between two states. 
We assume that in State A (the ON-state), the environment induces a 
certain concentration of activated factors, say on average $N_{TF}$ 
factors per cell (these may either be produced upon entering State A, or 
preexisting factors may be activated for binding by inducers that 
cause an allosteric transition). 
Let the growth rate or ``fitness'' of the organism in this state 
be $\phi_A$ if the factor is never bound (binding motif not present),
and $\phi_A + \delta \phi_A > \phi_A$ if the factor is always bound 
(see Table 1). 
When the environment is in State B (the OFF-state), 
let the fitness be $\phi_B$ if the factor is never bound and 
$\phi_B - \delta \phi_B < \phi_B$ if the factor is always bound. 
In the following we will assume that the concentration of activated 
factors in State B is practically zero, so that the operator is never 
occupied (hence the parameter $\delta \phi_B$ does not enter our model).  

An example for the general situation described above 
could be the binding of the {\it lac}-repressor to its operator
in the {\it lac}-operon of {\it E. coli}. In this case, the ON-state
would be the glucose-rich environment, and the OFF-state
would be the glucose-poor
and lactose-rich environment. $\phi_{A,B}$ would be the growth rate of
{\it E.~coli} in the two environments in the absence of the 
{\it lac}-repressor. $\delta \phi_A$ would
be the increment in fitness when the
unnecessary {\it lac}-operon is turned off, and $\phi_B - \delta\phi_B
\approx 0$ is the deleterious situation when lactose is present as the 
main source of sugar
but the {\it lac}-operon is not operative due to the 
undesirable binding of the repressor.

\begin{table}
\begin{tabular}{|l|c|c|}
\hline
 & State A & State B\\
\hline
factor unbound & $\phi_A$ & $\phi_B$\\
\hline
factor bound & $\phi_A + \delta \phi_A$ & $\phi_B - \delta \phi_B$\\
\hline
\end{tabular}
\caption{Fitness of the organism in two different cellular states, with
or without the binding of the transcription factor.}
\end{table}

In this study, we will discuss mainly the time-averaged effect over 
evolutionary time scales, which are much larger than the time scales of 
cellular or environmental fluctuations. 
We choose $\tau/\ln 2$ as our unit of time, with $\tau$ denoting the 
average generation time in the 
absence of the factor, so that the time averaged growth rate 
there can be set to 1. 
We assume that the cell can quickly adjust the cellular concentration of the 
factor\footnote{In the present article, we do not consider 
the `search problem' of how a transcription factor locates its operator 
among millions of other sites on the DNA 
(see \citet{berg81,winter81a,winter81b} for a thorough experimental 
and theoretical investigation of this problem, and 
Gerland, Moroz, and Hwa (2001, submitted for publication) for a 
bound on the protein-DNA interaction parameters that results from 
the requirement of reasonable search times). 
Rather we treat protein-DNA binding as an equilibrium process characterized 
only by the binding probability. 
This is justified by the fact that the search time is typically on the 
order of 1 min, which is much smaller than the characteristic time scale 
of changes in gene expression.}  
so that the operator with sequence $\vS$ is occupied with probability 
$P_{\vS}$ in the ON-state and never occupied in the OFF-state. 
It is then plausible to assume that the time averaged growth rate 
$\overline{\Phi}_{\vS}$ depends linearly on $P_{\vS}$ (see also the 
discussion in the section entitled `Selection threshold and fuzzy motifs'), 
\begin{equation}
  \label{fitness}
  \overline{\Phi}_{\vS} = 1 + \alpha \cdot P_{\vS}\;.
\end{equation} 
Here, $\alpha$ is a dimensionless parameter which characterizes 
the {\em selection pressure} on the binding sequence $\vS$. 
In the limit $\alpha\ll 1$, there is hardly any selection pressure on the 
sequence at all;
the opposite limit $\alpha\rightarrow\infty$ corresponds to the case where 
the failure of the factor-operator binding is lethal to the organism. 
If the fraction of time the bacteria population encounters environment A
is $\overline{f}$, the selection pressure becomes  
\begin{equation}
\label{alpha}
\alpha \equiv \overline{f} \cdot \delta\phi_A\;.
\end{equation} 
In an experiment, $\alpha$ can be adjusted according 
to Eq.~(\ref{alpha}) by changing the fraction of time $\overline{f}$ the
ON-state is presented. Below, we will investigate the statistics 
of the selected sequence $\vS$ for a wide range of $\alpha$'s.

\vspace{24pt}
\noindent{\bf Mutation process} 
\vspace{12pt}

We consider only single-nucleotide substitutions, and focus
on mutations in the binding sequence $\vS$, assuming that 
the net result of mutation and selection on the rest of the genome 
gives the overall background fitness of 1 (with our time unit 
of $\tau/\ln 2$). 
Furthermore, we neglect the difference between transversions and 
transitions, and assume a constant rate $\nu_0$ at which a base 
mutates into any other base. 
The total mutation rate of a site of length $L$ is then 
$\nu=\nu_0\,L$. 
For bacteria such as {\it E. Coli}, $\nu_0$ is on the order of 
$10^{-9}$ under normal conditions and hence $\nu\sim 10^{-8}$. 
The mutation rate is larger for RNA viruses
which rely on the less accurate reverse transcriptase for replication.
For that case, $\nu_0$ is in the range $10^{-5}$ to $10^{-4}$ and 
hence $\nu\sim 10^{-4}-10^{-3}$. 

\vspace{24pt}
\noindent{\bf Binding probability} 
\vspace{12pt}

\begin{figure}
\includegraphics[width=4.2cm]{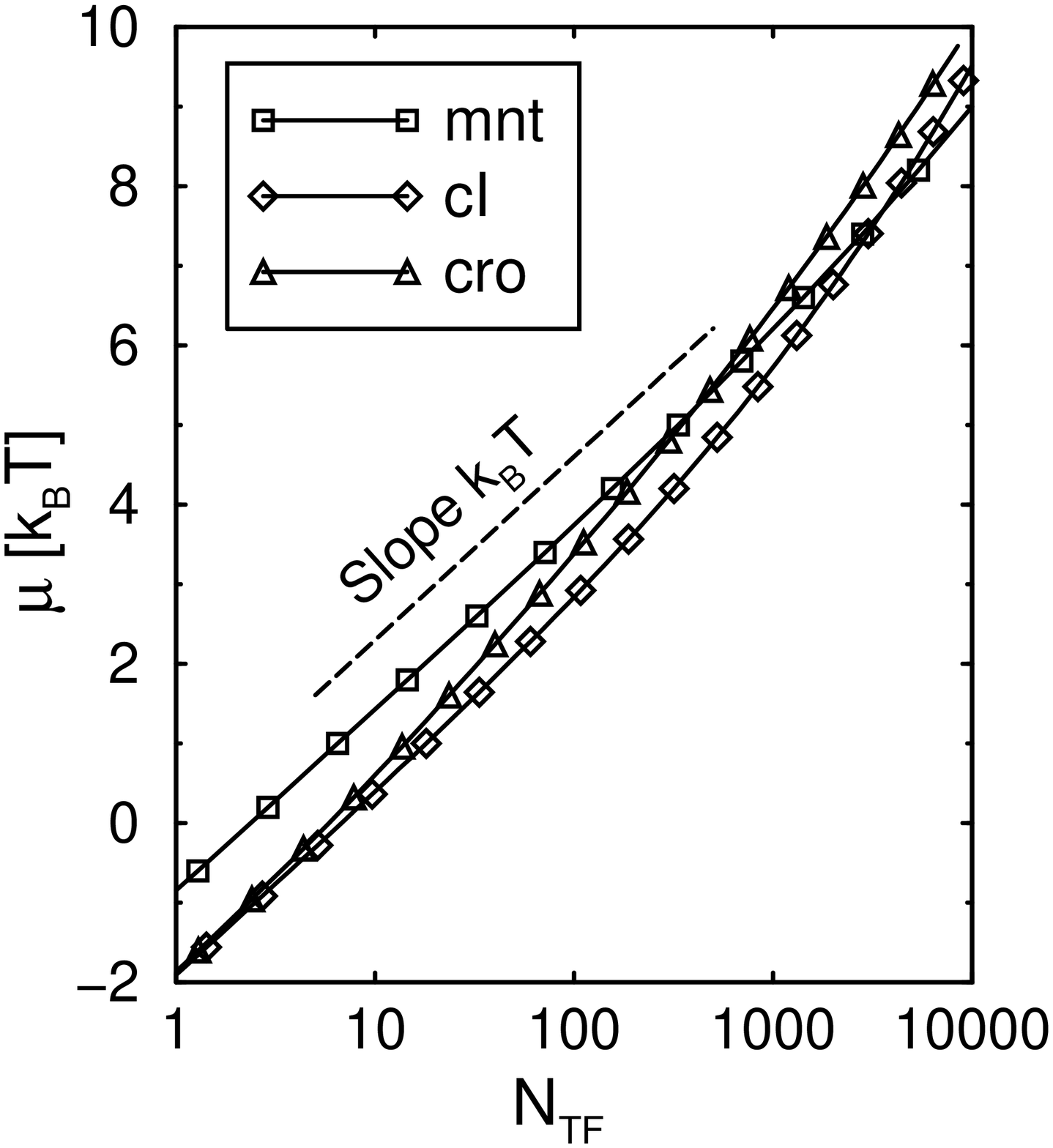}
\includegraphics[width=4.0cm]{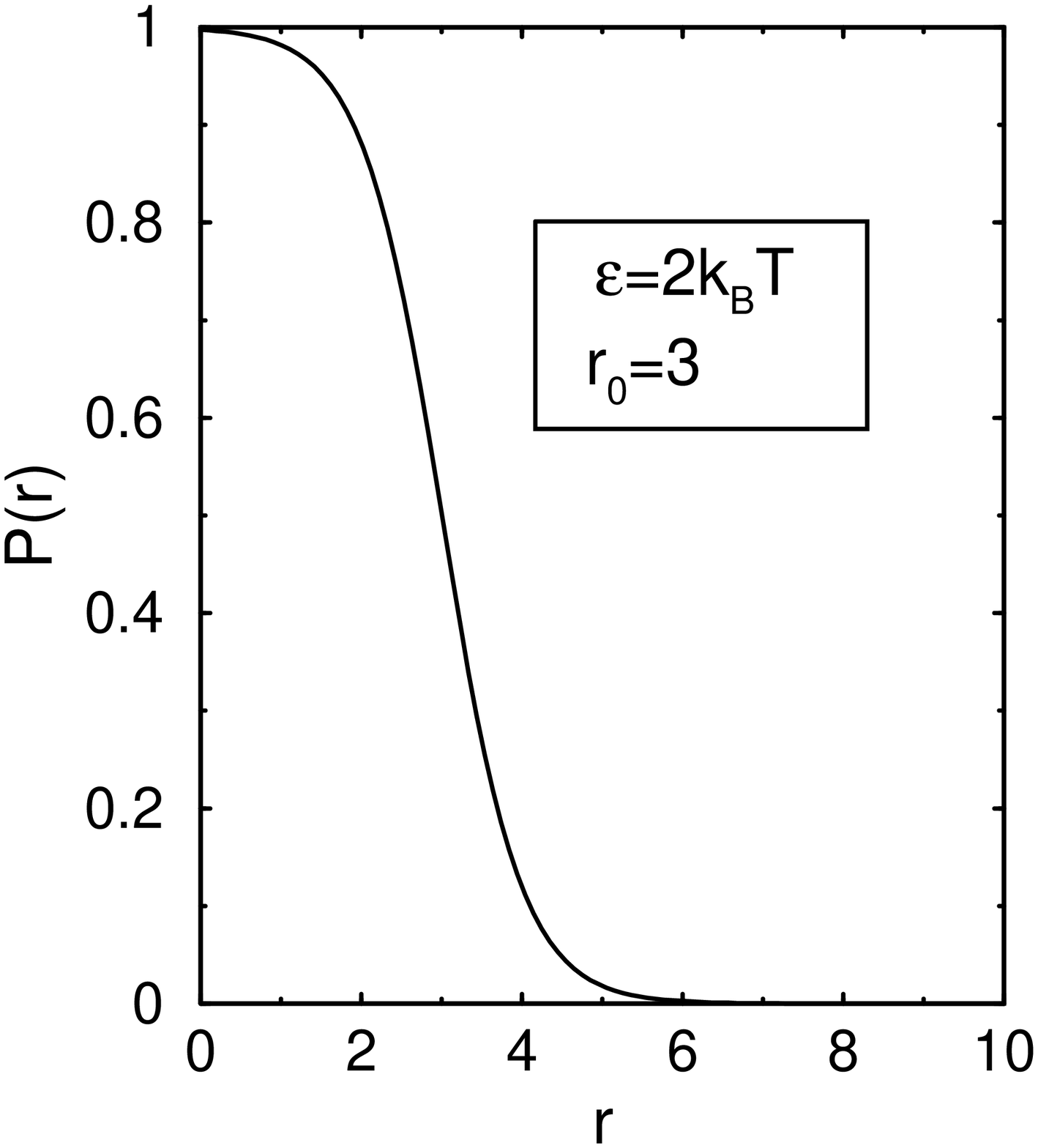}
\caption{\label{fig1} 
Left: chemical potential $\mu$ (in units of $k_BT\approx0.6$kcal/mol) 
as a function of the number of factors $N_{\rm TF}$ in the cell. 
The solid lines indicate the result of a numerical calculation using 
the Mnt energy matrix \citep{fields} on the Salmonella genome, 
the cro energy matrix \citep{takeda}, and the lambda repressor (cI) 
energy matrix \citep{sarai} on the {\it E. coli} genome. 
Here, we assumed  
that all of the factors in the cell are bound to DNA. 
The dashed line delineates the slope $k_BT$ which would be expected  
from the considerations of Gerland, Moroz, and Hwa 
(2001, submitted). 
Right: binding probability versus the number of mismatches from the best 
binding sequence, according to Eq.~(\ref{binding_prob}) with realistic 
parameters. 
}
\end{figure}

As mentioned above, the binding (free) energy $E_{\vS}$ of the transcription
factor to the binding sequence $\vS$ is given, to a good approximation, by 
a sum over independent contributions from each nucleotide at the $L$ 
significant positions~\citep{fields,stormo}, 
i.e. $E_{\vS} = \sum_{i=1}^L \eps_i(S_i)$.
Each of these positions typically prefers a 
particular nucleotide by a binding energy of several $k_BT$'s 
(we exclude from $\vS$ those positions which contribute only
a fraction of $k_BT$ towards the total binding energy). 
Furthermore we adopt the `two-state model'~\citep{vonhippel86,berg87}, 
by assuming that each $\eps_i(S_i)$ can only take on two possible values, 
0 if $S_i$ matches the preferred base $S_i^*$, or $\eps>0$ for a mismatch, 
i.e. $\eps_i(S_i)=\eps\cdot (1-\delta_{S_i,S_i^*})$. 
The binding energy of a site $\vS$ is then only a function of the number 
of mismatches, or Hamming distance $r_{\vS}=|\vS-\vS^*|$, from the optimal 
binding sequence $\vS^*$, i.e.\footnote{Note that the approximate 
linear relationship between the binding energy and 
the number of mismatches, Eq.~(\ref{energy}), breaks down when $E$ reaches 
a certain non-specific binding energy $E_{ns}$ \citep{winter81b}.
Beyond this point, the binding energy remains constant. 
However, the expression (\ref{binding_prob}) nevertheless provides an
useful description of the binding probability  over the entire range 
$0\le r\le L$, since $P(r)$ is essentially zero when $E=E_{ns}$.},
\begin{equation}
  \label{energy}
  E_{\vS} = E(r_{\vS}) = r_{\vS}\,\eps\;.
\end{equation}
Given its binding energy, the average occupancy of a site is determined by 
equilibrium thermodynamics. 
Since a binding site can only be occupied or unoccupied (but not multiply 
occupied), its binding probability $P_{\vS}=P(r_{\vS})$ 
is given by a Fermi function,  
\begin{equation}
  \label{binding_prob}
  P(r)=\frac{1}{1+e^{\eps(r-r_0)/k_BT}}\;, 
\end{equation}
which is also known as Arrhenius function, see, e.g., \citet{atkins}.
Here, $\mu = \eps\,r_0$ is the chemical potential for the transcription 
factors in the ON-state (this function is plotted in 
Fig.~\ref{fig1} (right) with realistic parameter values).
Note that $r_0$ corresponds to the number of mismatches for which the 
probability of binding is $50\%$.

In total, we are left with three dimensionless parameters for the 
two-state model of protein-DNA binding: $L$, $\eps/k_BT$, and $r_0$. 
As mentioned before, the number of significant positions in a binding site 
is typically in the range $10\le L\le 15$. 
By inspection of the known binding energies for exemplary transcription 
factors~\citep{fields,takeda,sarai,oda}, we find the mean specificity 
of the significant sites to be typically $\eps=1\sim3\,k_BT$. 
In (Gerland, Moroz, and Hwa, 2001, submitted for publication), it is 
argued on rather general ground that this is 
actually the optimal range of $\eps$ for the transcription factors. 
The chemical potential $\mu$ depends directly on the average number of 
factors $N_{\rm TF}$ in the cell;
the work of Gerland, Moroz, and Hwa (2001, submitted for publication)
suggests that it can be approximated by 
$\mu \approx \mu_0 + k_B T \ln [N_{\rm TF}]$, 
where $\mu_0$ represents the binding free energy of a single factor to 
the rest of the genome. 
For those factors whose binding energies $\eps_i(S_i)$ have been measured, 
we find $\mu_0\approx -0.8 k_BT$ (mnt) and 
$\mu_0\approx -1.9 k_BT$ (lambda-repressor and cro).
Hence, $\mu\approx k_BT \ln[N_{\rm TF}]$; see Fig.~\ref{fig1} 
for details. 
For $\eps =2k_BT$ and $N_{\rm TF} = 50 \sim 5000$, 
we get $r_0 = \mu/\eps = 2 \sim 4.3$.
Clearly, $r_0$ is the parameter that we have the least information about;
but we  see that it has a limited range, and in any case, 
most of our qualitative conclusions will be insensitive to 
the specific choice of $r_0$. 
[Note that the above analysis is for factors that have a binding site 
only in a single regulatory region. 
For those factors which are {\em global} regulators and 
have many operators located throughout the genome 
(e.g., the factor CRP in {\it E.~coli}), the number 
$N_{\rm TF}$ above needs to be appropriately adjusted 
by the number of operators \citep{sengupta}.]

\vspace{24pt}
\noindent{\bf Evolution equation} 
\vspace{12pt}

In this study, we will focus on the {\em steady-state} properties of the 
mutation/selection process defined above.
For large population size and close to the steady-state, we may consider 
only the dynamics of the {\em average} population and neglect fluctuations 
due to the discreteness of the individual organisms.
We denote the average number of individuals at time $t$ with binding 
sequence $\vS$ by $N_{\vS}(t)$. 
The time evolution of $N_{\vS}(t)$ is described by 
\begin{eqnarray}
  \label{evol_equation}
  \frac{\partial}{\partial t}\,N_{\vS}(t) &=& 
  \frac{\nu_0}{\cA-1} \sum_{\vS'} N_{\vS'}(t)
  \,\delta_{|\vS-\vS'|,1}+ \nonumber\\
  & & {}-\nu\,N_{\vS}(t) + \overline{\Phi}_{\vS}\,N_{\vS}(t)\,.
\end{eqnarray}
The first term on the right hand side describes the mutational flow into 
$N_{\vS}$ from all sequences $\vS'$ that are a single-nucleotide mutation 
away, while the second describes the reverse process. 
The third term represents the (time-averaged) 
selection/amplification process. 
Eq.~(\ref{evol_equation}) is similar to the ``para-muse model'' 
considered in a different context by \citet{baake}. 

Since the fitness function (\ref{fitness}) depends on the sequence $\vS$ 
only through the binding probability $P_{\vS}$, which
depends only on the number of mismatches $r$ according to 
Eq.~(\ref{binding_prob}), it is advantageous to 
introduce a `radial distribution' $N(r,t)$ in the (discrete) Hamming 
distance space \citep{nowak},  
\begin{equation}
  \label{Nradial}
  N(r,t)=\sum_{\vS} N_{\vS}(t)\,\delta_{r,\,|\vS-\vS^*|}\;. 
\end{equation} 
With $\bph(r_{\vS})\equiv\overline{\Phi}_{\vS}$ denoting the 
`radial fitness' function, the evolution equation 
for $N(r,t)$ becomes 
\begin{eqnarray}
  \label{discrete_evolution}
  \frac{\partial}{\partial t}N(r,t) &=& \bph(r) N(r,t) + 
  \frac{\nu_0}{\cA\!-\!1}\Delta_r\Big[(r\!+\!1)N(r\!+\!1,t)\Big]\nonumber\\
& & \quad {}-\nu_0\,\Delta_r \Big[(L\!-\!r)N(r,t)\Big]\;,
\end{eqnarray}
where $\Delta_r[f(r)] \equiv f(r)-f(r-1)$ denotes the discrete derivative,
and 
\begin{equation}
\label{mesa}
\bph(r) = 1 + \alpha P(r)
\end{equation}
is a mesa-shaped fitness landscape.
Eq.~(\ref{discrete_evolution}) 
is obtained by observing that there are $(L-r)(\cA-1)$ ways to mutate 
a site with $r$ mismatches into a site with $r\!+\!1$ mismatches, 
$r$ ways to mutate it into a site with $r\!-\!1$ mismatches, and 
$r(\cA-2)$ ways to mutate a site without changing the number of 
mismatches. 

We will characterize the predictions of our model by numerically 
integrating the discrete radial evolution equation (\ref{discrete_evolution}) 
using the set of realistic parameters given above. 
However, to gain insight about the qualitative behavior of the model, we also 
analyze the continuum-space evolution equation obtained in the limit of 
large $L$, 
\begin{eqnarray}
  \label{general_evol}
  \frac{\partial}{\partial t}n(r,t) &=& 
  \frac{\partial}{\partial r}\left[D(r)\,\frac{\partial}{\partial r}
  \,n(r,t)-v(r)\,n(r,t)\right] \nonumber\\ & & 
  {}+\overline{\varphi}(r)\,n(r,t)\;,
\end{eqnarray} 
where we used $n(r,t)$ and $\overline{\varphi}(r)$ to denote the continuum 
generalization of the functions $N(r,t)$ and $\bph(r)$ respectively.
Note that the mutational dynamics is {\em locally conservative}, with a 
local current $j(r,t) = D(r)\,\partial_r n(r,t)-v(r)\,n(r,t)$. 
The appropriate boundary conditions are $j(0,t)=0$ and $j(L,t)=0$.

The continuous radial evolution equation (\ref{general_evol}) reduces the 
evolutionary dynamics to a simple one-dimensional drift-diffusion equation, 
where the `diffusion coefficient' $D(r)$ and the `drift velocity' $v(r)$ 
are explicitly given by 
\begin{eqnarray}
  \label{D}
  D(r) &=& \frac{\nu}{2} \left(1-\frac{\cA\!-\!2}{\cA\!-\!1}
  \,\frac{r}{L}\right)\;,\\
  \label{v}
  v(r) &=& \nu \left(1-\frac{\cA}{\cA\!-\!1}\,\frac{r}{L}\right)\;.
\end{eqnarray} 
Note that both $D$ and $v$ are proportional to the overall mutation 
rate $\nu=\nu_0 L$, with only weak dependence on $r$ for $r\ll L$. 
The drift velocity drives the distribution away from the optimal binding 
site at $r=0$, simply because the number of sequences with 
a fixed number of mismatches $r$ increases quickly with $r$. 
This purely entropic bias changes sign at $(\cA\!-\!1)L/\cA$, which is 
the average number of mismatches in a random binding sequence. 
Also note that the $r$-dependence of the diffusion coefficient disappears 
when $\cA=2$, because for a two-letter alphabet, every mutation implies a 
change in $r$. 
For $\cA>2$, there are mutations which do not change the Hamming distance 
and hence do not affect the diffusion process. 
This effect is reflected in the reduction of $D$ in Eq.~(\ref{D}).

Our continuous radial evolution equation (\ref{general_evol}) is somewhat 
reminiscent of the evolution equation in fitness space introduced by 
Tsimring, Levine, and Kessler \citep{herbie} in a general population 
genetics context. 
However, with our concrete model for protein-DNA binding, we can work 
directly in genotype space, which will enable us to make explicit 
predictions on the behavior of the binding sites. 
\section{Selection threshold and fuzzy motifs}
In this section, we use the evolutionary model (\ref{discrete_evolution})
described in `Model and Equations' to address the following questions:
How large a selection pressure is needed for the maintenance of binding 
motifs? And can the fuzziness of the motifs be accounted for by the 
balance between mutation and selection? We will first provide an analytical
solution to a simplified continuum model, and then show by numerical 
simulation that the qualitative features of the solution hold even for
small system such as $L=10$. 
We will compare these results to available data and discuss 
experimental ramifications.

\vspace{24pt}
\noindent{\bf Analytical results} 
\vspace{12pt}

Various properties of the continuum model specified by 
Eqs.~(\ref{general_evol}-\ref{v}) can be obtained 
exactly\footnote{Inspired by the present system,
solution of the mean-field evolution model for general mesa-like 
fitness landscape has recently been developed by \citet{peliti}.}. 
Here, we will present the results and discuss various
qualitative features of the solution, in particular,
the existence of a critical selection pressure for the maintenance of
the binding motifs. Even though the continuum model is meaningful only 
for $L \gg r_0 \gg 1$, we will see from numerical simulation  
that the qualitative features are valid
even for the more realistic parameter range where $r_0$ is not much 
larger than one, and $L \sim 10$. 

For the analytical study, we neglect the $r$-dependence of the diffusion 
coefficient (\ref{D}) and the drift velocity (\ref{v}), and use 
$D=\nu/2$, $v=\nu$. 
This is justified as long as $r_0\ll L$ since as we will see, most of the 
interesting ``action'' of this system occurs around $r=r_0$. 
Eq.~(\ref{general_evol}) then reduces to the asymmetric ``quantum
well'' problem well studied in the context of various statistical 
mechanics problems \citep{hatano}. 
[It differs from the DNA unzipping problem studied by 
\citet{lubensky} only by an (unimportant) boundary 
condition at $r=0$.] 
An explicit solution can be obtained by
further approximating the Fermi function (\ref{binding_prob}) by the 
Heavyside step function $\theta(r)$, such that the fitness landscape becomes
\begin{equation}
\label{step}
\bph(r) = 1 + \alpha \, \theta(r_0-r).
\end{equation}
This idealized form of the fitness function is known as 
{\em truncation selection} \citep{kondrashov}.

The solution of this simplified continuum problem
is of the form $n(r,t) = n_0(r) e^{\gamma t}$, 
where $n_0(r)$ is the stationary distribution associated with the
largest growth rate $\gamma$. 
It is controlled by one dimensionless parameter, the effective 
selection pressure
\begin{equation}
\label{talpha}
\talpha \equiv \frac{2\alpha}{\nu}.
\end{equation}
We have $\gamma = 1$ if $\talpha$ is below the critical value
\begin{equation}
  \label{alpha_c}
  \talpha_c = 1 + \frac{\eta_c^2}{r_0^2}\;,
\end{equation}
where $\eta_c$ is of order one and depends only 
weakly\footnote{For $1 \ll r_0 \ll L$, $\eta_c$ is given to a good 
approximation by the the solution of the equation 
$\eta_c = - r_0 \cdot \tan(\eta_c)$ and hence 
$\eta_c \in [\frac{\pi}{2},\pi]$ .} on $r_0$.
In this regime, $n_0(r)$ is given by the continuum version of the
(skewed) binomial distribution
\begin{equation}
\label{binomial}
\Omega(r) =  (\cA-1)^r \left({L \atop r}\right)
\end{equation}
as if the fitness plateau at $r<r_0$ is not present. 
For $\talpha > \talpha_c$, the solution is given 
in terms of the eigenvalue problem
\begin{equation}
  \label{fx}
  y''(r) + \talpha\, \theta(r_0-r)\,y(r) = \lambda\,y(r)
\end{equation}
with the boundary condition $y(0)=y'(0)$, where $y(r)$ is the 
eigenfunction corresponding to the largest eigenvalue, 
$\lambda(\talpha)$, which must exceed $1$.
[Thus, the precise definition of $\talpha_c$ is $\lambda(\talpha_c)=1$.]
In this regime, the growth rate becomes  
$\gamma = 1 + (\lambda - 1) \nu/2 \ge \bph_0$, 
and the stationary distribution is $n_0(r) = y(r) e^r$.
The form of the latter can be straightforwardly analyzed for 
large $\talpha$'s (such that $\lambda > 1$). It is
strongly peaked at $r^* \lesssim r_0$, indicating that the motifs
are {\em marginally conserved} or {\em maximally fuzzy}\footnote{In the 
context of protein folding, it has been pointed out by R.\ Goldstein 
that the balance of mutation and selection may lead to maximal fuzziness
in the space of amino acid sequences~\citep{goldstein}.
Our results are similar in spirit, but more explicit due to the simplicity
of the protein-DNA binding.} above the selection threshold.

A {\em phase transition} occurs at $\talpha=\talpha_c$ 
where the stationary distribution switches 
from being mostly confined in the region $r<r_0$ (localized phase) to the
binomial distribution (delocalized phase)  when $\lambda$ approaches $1$.
This phase transition belongs to the same class of transitions as the 
one described by Eigen in the context of quasi-species evolution 
\citep{eigen71,eigen89,higgs}. 
The critical selection pressure $\alpha_c\sim\nu_0\cdot L$ is 
recognized as the well-known form of the ``error threshold''. 
Note also the dependence of $\alpha_c$ on $r_0$ as given in 
Eq.~(\ref{alpha_c}): $\alpha_c$ decreases upon increasing $r_0$, 
and since $\lambda$ is a monotonously increasing function of 
$\alpha-\alpha_c$, the effective growth rate $\gamma$ will also 
increase.  
This implies that a wider fitness landscape has a smaller mutational 
load and a larger effective fitness, which is a known result, see, 
e.g. \citet{schuster}. 

The ``order parameter'' of the phase transition is the average
number of mismatches in the stationary state, 
 $\br=\int_r\, r\, n_0(r)/\int_r\,n_0(r)$.
In the localized phase, $\br\simeq r_0$, while 
$\br=(\cA-1)L/\cA\rightarrow\infty$ in the delocalized phase. 
When $\talpha$ approaches $\talpha_c$ from above, $\br$ diverges as
\begin{equation}
  \label{order}
  \br \propto (\talpha-\talpha_c)^{-1} \qquad {\rm for} 
  \qquad \talpha \gtrsim \talpha_c\;. 
\end{equation}
indicating that this is a {\em second-order} phase
transition\footnote{It should be noted that both the critical value
$\talpha_c$ and the divergence of $\br$ near $\talpha_c$ are modified
if one explicitly includes the {\em time dependence} of the fitness
landscape. In particular, if we take the fitness to be $\phi(r,t)
\propto f(t) P(r)$ ($f(t)=1$ in the ON-state and $f(t)=0$ in the 
OFF-state), with a {\em stochastic}
$f(t)$, then the evolution dynamics becomes equivalent to the class of 
time-dependent depinning problems studied by \citet{lubensky}, 
with the critical behavior $\br \propto (\talpha-\talpha_c)^{-2}$.}.

\vspace{24pt}
\noindent{\bf Numerical results} 
\vspace{12pt}

\begin{figure}
\includegraphics[width=8cm]{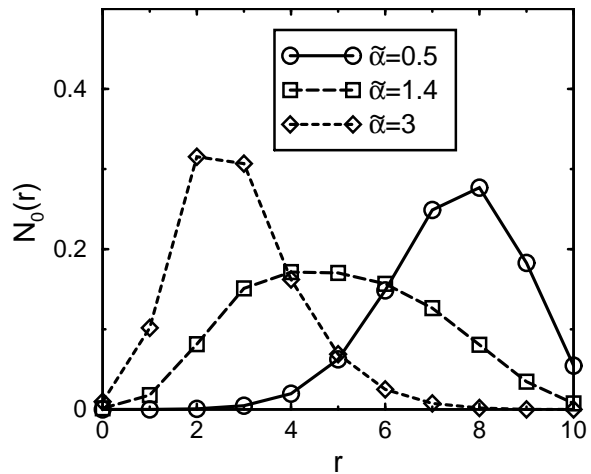}
\caption{
The stationary distribution function $N_0(r)$ for the discrete 
model with $L=10$, $r_0=3$, and  $\eps=2k_BT$. 
At $\talpha=0.5$, the distribution is indistinguishable from 
$\Omega(r)$.}
\label{fig2}
\end{figure}

In order to test whether the behavior derived above for the 
simplified  continuum model holds approximately
also for the discrete model (\ref{discrete_evolution}) with realistic 
parameters, we performed a number of numerical studies.
We determined the steady-state distribution $N_0(r)$ of 
Eq.~(\ref{discrete_evolution}) over a range of values of
$\talpha$ for two sets of parameters: 
(a) a nearly-continuum model, 
with $L=1000, r_0=30$, and the step function landscape (\ref{step});
and (b) the discrete model with $L=10, r_0=3$, and the Fermi function 
landscape (\ref{mesa}) with $\eps=2k_BT$.
Fig.~\ref{fig2} shows the stationary distribution $N_0(r)$ for the 
discrete model in the delocalized regime ($\talpha=0.5$), 
in the localized regime ($\talpha=3.0$), and in the 
crossover\footnote{With a small system such as $L=10$, the 
critical point of the phase transition becomes a crossover region, in 
which the behavior of the stationary state changes smoothly.} region 
in between ($\talpha=1.4$). 
We see that $N_0(r)$ is peaked slightly below
$r_0=3$ in the localized regime, and becomes indistinguishable from the
binomial distribution (\ref{binomial}) in the delocalized phase. 
Note that the distribution is {\em broad} in the crossover regime, which 
is consistent with the finding of a continuous second-order transition 
in the continuum model (see the last section). 

\begin{figure}
\includegraphics[width=8cm]{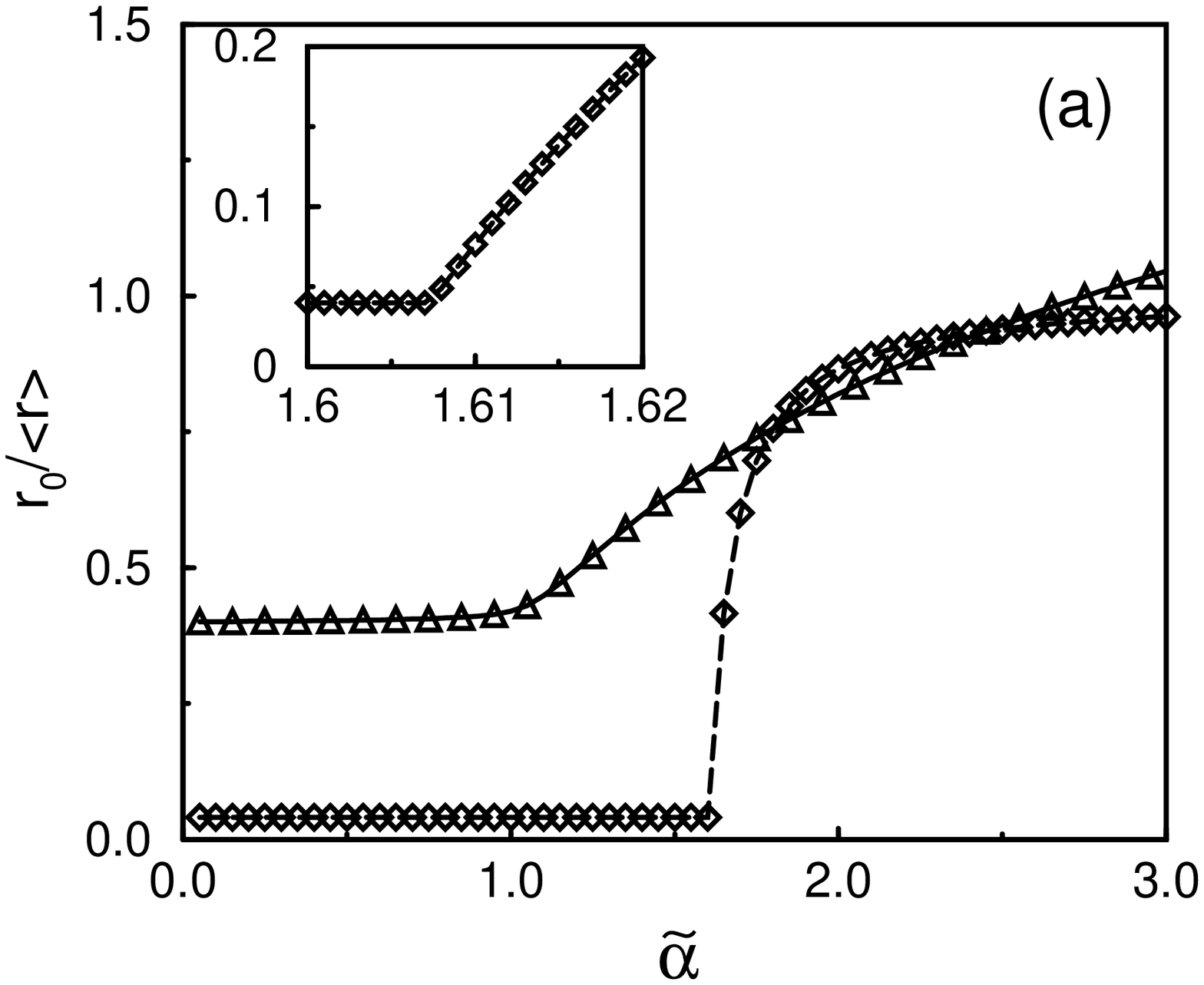}\\
\includegraphics[width=7cm]{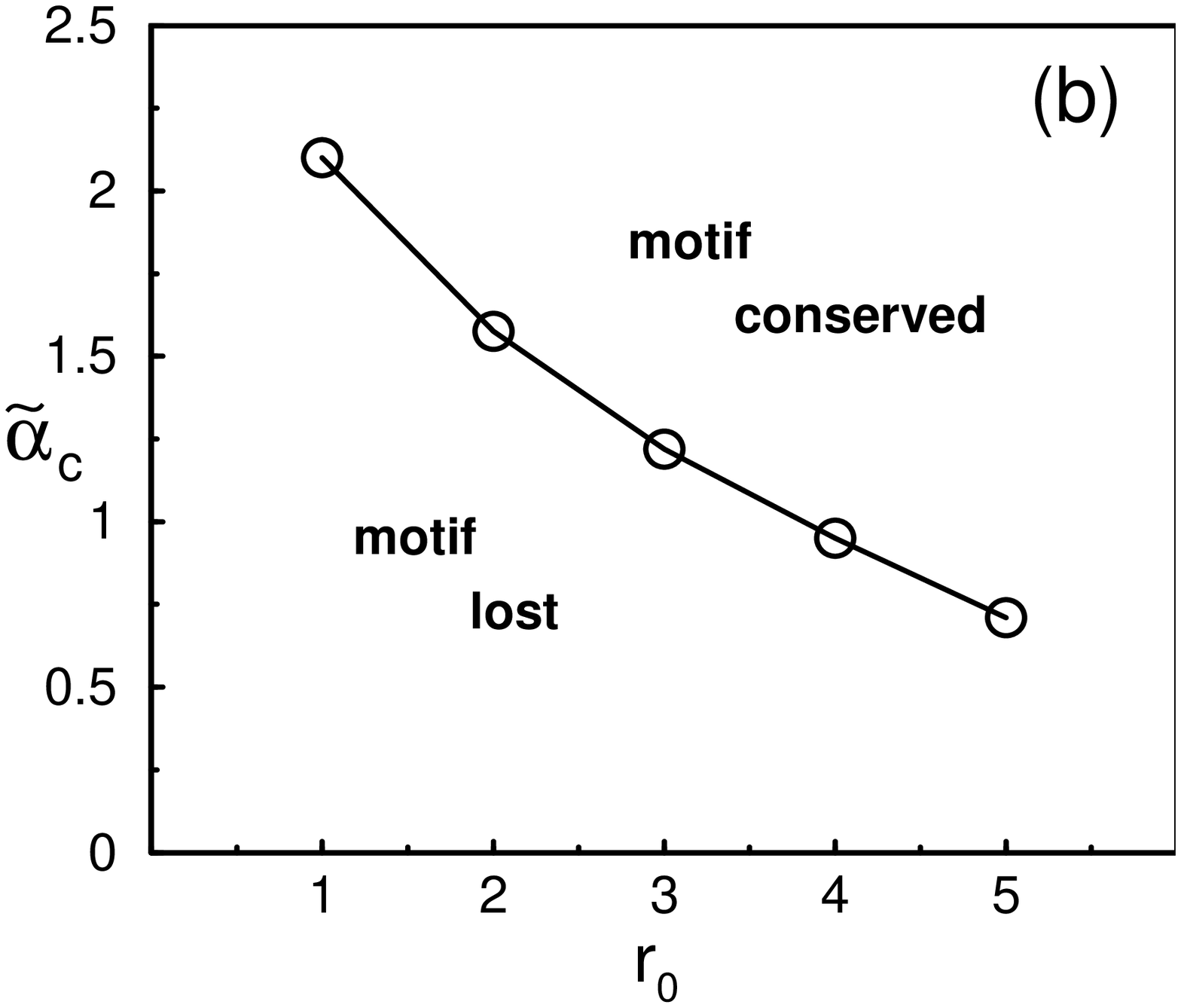}
\caption{
(a) The order parameters $r_0/\br$ for the nearly-continuum model 
(diamonds) and discrete model (triangles). The transition is
quite pronounced even for $L=10$. The inset shows a magnification of 
the critical region for the nearly-continuum model, which confirms the 
critical exponent of one in Eq.~(\ref{order}).
(b) The dependence of the threshold $\talpha_c$ on $r_0$ for the 
discrete model. The result $\talpha_c \sim 1$ is in qualitative
agreement with the formula (\ref{alpha_c}) derived for the continuum model.
}
\label{fig3}
\end{figure}

To make the comparison more quantitative,  we next examine the order
parameter $\br$.
Fig.~\ref{fig3}(a) shows $r_0/\br$ plotted as a function of $\talpha$, for 
the discrete model (triangles) and the nearly-continuum model (diamonds).
The nearly-continuum model displays a sharp transition at\footnote{The 
critical value for the nearly-continuum model deviates somewhat from the 
analytical result (\ref{alpha_c}). This deviation is due to the 
$r$-dependence of $v$ and $D$, which we neglected in the analytical 
solution of the continuum model.} $\talpha_c\approx 1.6$. 
The sharp transition becomes a pronounced crossover for the discrete 
model, but still with a relatively well-defined threshold $\talpha_c$. 
The $r_0$-dependence of $\talpha_c$ is plotted in Fig.~\ref{fig3}(b) 
over the relevant interval $1<r_0<5$ 
(here, we have defined the threshold $\talpha_c$ as the value of $\talpha$ 
where the derivative of $r_0/\br(\talpha)$ is maximal). 
We see that it is relatively insensitive to the precise value of $r_0$, 
with $\talpha_c \sim 1$ as given qualitatively by the formula (\ref{alpha_c}).

\vspace{24pt}
\noindent{\bf Viral and bacterial evolution} 
\vspace{12pt}

We expect the selection threshold described above to be detectable in  
evolution experiments with RNA viruses. 
The total mutation rate $\nu$ for the binding 
site for RNA viruses is of the order $10^{-3} \sim 10^{-4}$ for a binding
sequence of length $L=10$. 
Assuming that the fitness gain of the virus in the ON-state 
(i.e., the factor $\delta \phi_A$ in Eq.~(\ref{alpha})) is of the 
order of $1\% \sim 10\%$, then the effective selection pressure $\talpha
= 2\overline{f} \delta\phi_A/\nu$ on the viral regulatory sequence
becomes of the order $\talpha_c \sim 
O(1)$ if the fractional exposure $\overline{f}$ 
to the ON-state is set at a few percent level. By varying $\overline{f}$
over the range of several percents, we expect that the phase transition
should be observable. Moreover, the anomalous dependence (see footnote 4) 
of the selection threshold
on the temporal variation $f(t)$ should also be observable by applying 
controlled temporal changes to the environment.
The stationary distribution $N_0(r)$ itself can be monitored
in principle by sequencing a reasonable number (say $100$) 
viral regulatory sequences after stationarity is reached.

\begin{figure}
\includegraphics[width=6cm]{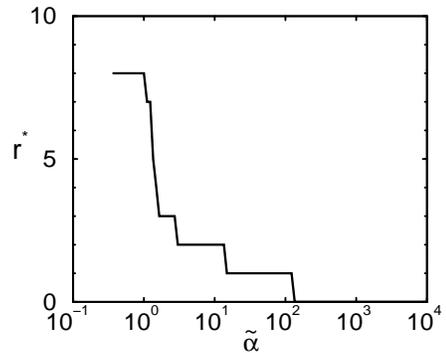}
\caption{\label{fig4}
The peak position $r^*$ of the stationary distribution $n_0(r)$  
for the Fermi function landscape (\ref{mesa}) with
$L=10$, $r_0=3$, and $\eps=2k_BT$.}
\end{figure}

A very different situation is expected for the evolution of bacteria
or even DNA viruses. The total mutation rate $\nu$
is of the order $10^{-8}$ for bacteria and $10^{-6}$ for DNA viruses. 
Consequently, $\talpha$ is expected to be
 four  orders of magnitude larger than $\talpha_c$ for bacteria
and two orders larger for DNA viruses. What is the behavior
of the discrete model at such large values of $\talpha$ ?
In Fig.~\ref{fig4}, we show the position of the peak $r^*$ of the
distribution $N_0(r)$ obtained numerically as a function of 
$\talpha$ on a logarithmic scale. 
For values of $\talpha$ exceeding $\sim 140$, we find the peak
is pushed down to $r^*\!=\!0$, contrary to the fuzziness depicted in
Fig.~\ref{fig3}.

\begin{figure}
\includegraphics[width=6cm]{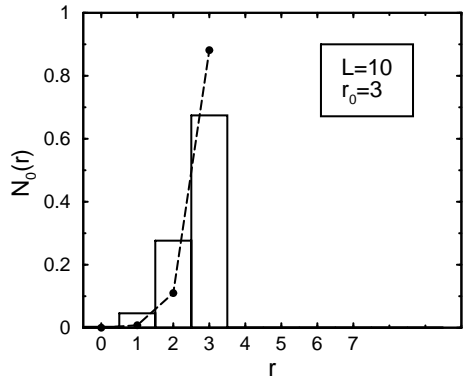}
\caption{\label{fig5}
The stationary distribution $N_0(r)$ for the infinite-mesa landscape
(\ref{infinite}) with $r_0=3$ (histogram). 
For comparison the density of states, i.e. the binomial distribution 
(\ref{binomial}), cut off at $r=r_0$, is also shown (dashed line).
}
\end{figure}

This behavior is obviously an artifact of the specific feature of the 
Fermi function landscape used in (\ref{mesa}): for very large
$\talpha$'s, there is an incentive for the distribution to move towards
small $r$'s due to the very slight increase in the value of $P(r)$
for smaller $r$'s. But it is unreasonable to expect that the simple relation
between the binding probability $P(r)$ and the fitness function $\bph(r)$
assumed in this study to hold down to very small differences in $P(r)$.
Aside from various kinetic effects of binding and temporal variations 
of the environment that we have neglected, 
fluctuations due to finite population size (e.g., genetic drift) 
simply do not allow for the population to resolve the very small differences
in fitness due to the small differences in $P(r)$; see the theory of 
nearly-neutral evolution \citep{ohta}. 
Thus, $\bph(r)$ should be effectively $r$-independent for small $r$'s.
This can be implemented by replacing $\bph(r)$ by a constant value
$\bph(r_0')$ when $\bph(r_0'-1)-\bph(r_0')$ is below some resolution limit
(set by the effective population size of the organism.)\footnote{This
modification of the fitness function should in principle also be applied
to the case of RNA virus evolution in the vicinity of the phase transition. 
However, it wouldn't make much of a difference there because the 
distribution would be already peaked away from small $r$.}
For low mutation rate (or large $\talpha$'s), 
this amounts to replacing the fitness function by an infinte square well:
\begin{equation}
\label{infinite}
\bph(r) = \left\{ 
\begin{array}{ll}
\alpha\to\infty& \mbox{if $r \le r_0$}\\
0 & \mbox{if $r>r_0$}
\end{array}
\right.\,.
\end{equation}
The stationary distribution obtained in this case depends only on the
width of the well $r_0$, and is shown in Fig.~\ref{fig5} for $r_0=3$. 
Note that it is highly peaked at $r_0$ as expected. 
Hence the binding sequence is fuzzy even as  $\talpha \to \infty$. 
However, it is different from simply truncating the 
binomial distribution (\ref{binomial}) for $r>r_0$ due to the 
mutational load, i.e., a fraction of the population
with $r=r_0$ will receive an additional deleterious mutation and not
make it to the next generation.

\vspace{24pt}
\noindent{\bf Comparison to known sites} 
\vspace{12pt}

It is useful to compare the above solution to biological data.
Unfortunately, polymorphisms in the same binding sequence 
across different strains
of a bacteria species are not yet readily available. Instead, we assume that
the different binding sequences (of the same transcription factor)
located in different regulatory regions across the genome
may be viewed as a sample of the 
stationary binding sequence distribution. This is clearly not the case
if the selection pressure is small, since close to the selection threshold,
even small differences in selection pressure experienced by the different 
binding sequences will produce different binding sequence distributions;
see Fig.~\ref{fig2}. But this should not be 
a concern for bacteria since $\talpha \gg \talpha_c$ there.
An obvious candidate is the binding sequences for the 
the well-known {\it E.~coli} global regulator CRP (also known as the 
catabolite activator protein, CAP), which is 
activated under low cellular glucose level \citep{saier}.
There are over one hundred CRP sites in the {\it E.~coli} genome.
We take from the RegulonDB database \citep{regulonDB} a list of 28 sites 
which are biologically confirmed binding sites and appear only once 
in a given regulatory region. (The case of 
multiple binding sites is discussed in the following section.)
The drawback of using CRP sites is that CRP is hardly ever the 
only regulator in a target regulatory region, and interaction with
other transcription factors can complicate the situation.

\begin{figure}
\includegraphics[width=6cm]{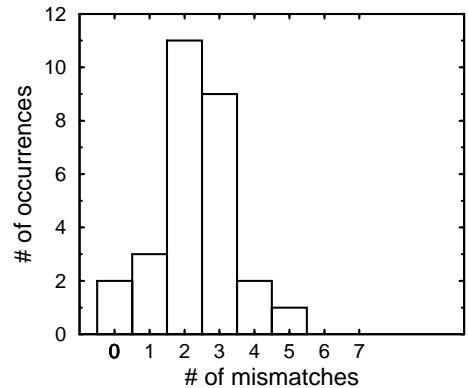}
\caption{\label{fig6}
Histogram of mismatches of known CRP (singlet) sites
from the consensus sequence {\tt TGTGA......TCACA}.
}
\end{figure}

Fig.~\ref{fig6} shows the histogram of the number of mismatches
of these CRP sequences from the consensus sequence
{\tt TGTGA......TCACA}. While it peaks at $r^* = 2 \sim 3$ similar to
the corresponding distribution 
of the infinite-mesa model in Fig.~\ref{fig5}, it is clear that the
distribution of the CRP sites is broader. The few outliers
at $r=0, 1, 4, 5$ may well be due to direct or subtle interaction 
with other factors which we have not considered in this simple model.
The existence of nearly equal peaks at $r=2$ and $r=3$ is more perplexing:
According to our model, the distribution should be peaked at the 
largest possible $r$. [For the $L=10$ sequence, entropy favors $r=3$
over $r=2$ by a factor of eight.] One possible cause of the discrepancy
may be the deviation of the real binding energy matrix $\eps_i(S)$ 
from the two state model. For instance, suppose the chemical potential $\mu$
in the ON-state is such that $r_0=\mu/\eps$ is slightly below $3$. Then
according to the pure two-state model, the maximum number of allowed
mismatches is $2$. 
However, small deviations in the binding energies from $\eps$
will allow a maximum of 3 mismatches in a subset of the $L$ positions,
thereby producing a distribution peaked at both
$r=2$ and $r=3$ as shown in Fig.~\ref{fig6}.
The actual stationary distribution of $r$ can be easily computed numerically
if the energy matrix is known. However at present, the authors know of
no example of a transcription factor whose binding energy matrix 
is measured and a large number of binding sequences are available.

A very different explanation of the data in Fig.~\ref{fig6} is the 
differential selection of each of the CRP motifs as alluded already
in ``Introduction''. Specifically, one can envision a situation where
the single ``ON-state'' assumption we adopted is not valid, and instead
the cell coordinates a {\em graded} response to cellular glucose shortage,
requiring different operons to turn ON at different (activated) CRP 
concentrations. [In this case, our assumption that the fitness function 
(\ref{fitness}) has a simple linear dependence on the binding probability 
obviously breaks down.] 
Within this scenario, the distribution in Fig.~\ref{fig6}
is solely a result of the functional need of the cell, and its
resemblance to the statistical distribution of Fig.~\ref{fig5}
would be fortuituous. Distinguishing between the different plausibilities
is important and can be done by either sequencing the CRP binding sites 
in a variety of related strains to accumulate statistics on polymorphism
for each site, or perform site-directed mutagensis to specific binding 
motifs and directly measure the fitness function.
In general, one may expect to find that the form of the fitness function 
depends on the biological function of the binding site. 
In particular, the form (\ref{fitness}) seems more likely to be appropriate 
for the case of transcriptional repressors than for activators, since 
repressors need to have a binding probability close to one, in order to 
efficiently suppress transcription from the promoter, which 
is active in the absence of the repressor. In the case of activators, the 
promoter has a very low basal level of transcription and even an 
activator with a relatively low binding probability can lead to a large 
effect on the transcription level. 
\section{Multiple binding sites}
It is well known that regulatory binding motifs often occur in doublets 
or even higher multiplets. 
For instance, the regulatory regions of the {\it E. coli} genes  
crp, dadA, \mbox{dsdXA}, fixA, glpFK, glpTQ, lac, manX, nagE, and tsx  
are some of the many regions that contain two CRP binding sequences. 
Here, we extend our model to account for the possibility of multiple 
sites that bind the same protein and regulate the same promoter. 
We will pursue the question of whether we can interpret regulatory regions 
with multiplets as being under higher selective pressure for factor 
binding than regulatory regions with single binding sites. 

Some factors (e.g., $\lambda$-repressor) bind {\em cooperatively} to 
binding sites, thereby effectively enhancing their DNA binding specificity. 
Cooperative factor binding can play an important and interesting role 
in transcription regulation (see, e.g., \citet{ptashnebook}), however 
it does so only for a fraction of the known multiplets, since many 
factors (such as CRP) have no binding domain for an attractive interaction 
between themselves. 
In the present study we exclude factor-factor interactions and explore 
possible selective advantages of multiple independent binding sites. 
This approach is similar in spirit to studies of gene duplication, which 
consider the evolution of multiple copies of the same gene 
(see, e.g., \citet{wagner}).  
One scenario could be that several bound transcription factors can 
simultaneously interact with polymerase to synergistically recruit 
(or repel) it more efficiently than a single factor would. 
For the case of CRP, this effect has been observed, and studied in 
detail, experimentally \citep{busby,langdon}. 
An individual organism with a multiplet of binding sites for a factor 
then has a fitness advantage over one with a single binding site, 
if a strong activation (or repression) is beneficial for the biological 
function. 
Consequently, selection would favor multiplets over singlets. 
On the other hand, random mutations tend to destroy the binding motifs, so 
that an additional motif is maintained only when its contribution to the 
fitness is sufficiently high. 
In the following, we explore this scenario within our model. 

Let us assume there are two binding sites in a certain regulatory region 
and ask whether they will be maintained by evolution. 
We begin by constructing a `two-site fitness function' that makes the 
selection mechanism outlined above explicit.  
As in the previous sections on the single-site problem, we assume that 
the state of the bacterium/virus alternates between an ON-state, where 
factor binding leads to a fitness gain, and an OFF-state, where factor 
binding has a negative effect. 
In the ON-state, let the fitness gain due to factor binding be 
$\delta \phi_{A1}$, $\delta \phi_{A2}$, or $\delta \phi_{A12}$, if a 
factor is bound to site 1 only, site 2 only, or both sites, respectively. 
Using the same arguments as for the single site case, the time-averaged 
fitness becomes 
\begin{equation}
  \label{2site_fitness}
  \overline{\Phi} = 1 + \alpha\cdot 
  [ P_1(1-P_2) + \sigma\,P_2(1-P_1) + \omega\,P_1P_2]\;,
\end{equation}
where $P_1$, $P_2$ denote the probabilities that a factor is bound to 
site 1, 2, which depend on the respective sequences (we neglect the 
possibility of cooperative binding at this point). 
The selection pressure, $\alpha$, has again the form 
(\ref{alpha}), with $\alpha=\overline{f} \, \delta\phi_{A1}$, 
while the `synergism factor' $\omega$ describes the fractional fitness 
advantage of two bound factors over just one, i.e. 
$\omega=\delta\phi_{A12}/\delta\phi_{A1}$. 
In the remaining term, the dimensionless coefficient $\sigma$ constitutes 
an `asymmetry factor' equal to the relative fitness gain 
$\delta\phi_{A2}/\delta\phi_{A1}$ (i.e. when $\sigma\ne 1$ one may 
distinguish between a more ``important'' and a less important site). 
Note that not only the selection pressure, but also $\omega$ and 
$\sigma$ may vary between different regulatory regions, even when they 
are controlled by the same factor, since both depend on the location of 
the binding sites with respect to the promoter and on the sequence of 
the promoter itself (see the discussion at the end of this section). 

As in the single site problem, we work in the two-state model 
approximation (see section `Binding probability'), so that the binding 
probabilities $P_1$ and $P_2$ only depend on the number of mismatches 
$r_1$ and $r_2$ in the respective site, and take the form 
(\ref{binding_prob}). 
When the selection pressure $\alpha$ is much larger than the mutation 
rate $\nu$ (as we typically expect in the case of bacterial evolution), 
we again invoke the argument that very small differences 
in the fitness function are hardly resolvable by finite populations, 
and therefore the fitness function should become neutral, i.e. 
$r$-independent, at small $r_1$ and $r_2$. 
This effectively amounts to using step functions for the binding 
probabilities, i.e. $P(r_{1,2})=1$ for $r_{1,2}\le r_0$ and 
$P(r_{1,2})=0$ for $r_{1,2}>r_0$. 
The two-site fitness function in $(r_1,r_2)$-space is then 
\begin{equation}
  \label{2Dfitness_bacteria}
  \overline{\phi}(r_1,r_2) = \left\{ 
  \begin{array}{ll}
    1+\alpha\; & \mbox{if $r_1\le r_0<r_2$}\\
    1+\alpha\sigma\; & \mbox{if $r_2\le r_0<r_1$}\\
    1+\alpha\omega\; & \mbox{if $r_1$, $r_2\le r_0$}\\
    1\; & \mbox{if $r_1$, $r_2>r_0$}
  \end{array}
  \right.\,.
\end{equation}
In order to simplify the discussion in the following, we will use the 
fitness function (\ref{2Dfitness_bacteria}) over the whole range of 
$\talpha$, since it yields a correct description at large $\talpha$, 
and at small $\talpha$, it produces no qualitative changes in the 
behavior of the stationary distribution compared to the smooth 
fitness function with $P(r)$ of the form (\ref{binding_prob}). 

\begin{figure}
\includegraphics[width=6.5cm]{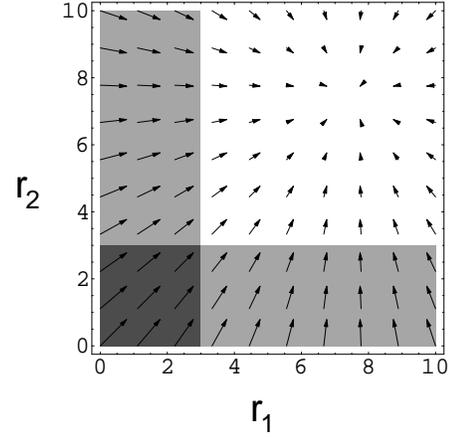}
\caption{\label{fig7}
Illustration of the drift-diffusion dynamics for the two-site problem. 
The arrows indicate the direction and magnitude of the drift velocity 
$\vec{v}=(v(r_1),v(r_2))$ in the continuum equation 
(\ref{2DcontinuumEq}), while the shading 
corresponds to the fitness function (dark means high fitness; here we 
used $\omega=2$ for the purpose of illustration, and $\sigma=1$)}
\end{figure}

It is straightforward to derive a two-site evolution equation 
analogous to Eq.~(\ref{discrete_evolution}), which describes the 
approach of the average distribution of mismatches $N(r_1,r_2,t)$ 
(neglecting fluctuation effects due to finite population size) 
to its stationary state $N_0(r_1,r_2)$. 
One obtains 
\begin{eqnarray}
  \label{2Ddiscrete_evolution}
  \frac{\partial}{\partial t}N(r_1,r_2,t) &=& \bph(r_1,r_2) N(r_1,r_2,t) + 
  \nonumber\\ & & \hspace*{-2cm} {}+\nu_0 \, \Delta_{r_1}
  \Big[(r_1\!+\!1)N(r_1\!+\!1,r_2,t) + \nonumber\\ & & \hspace*{-0.5cm} 
  {}-({\cal A}-1)(L-r_1) N(r_1, r_2,t)\Big]\nonumber\\ & & 
  \hspace*{-2cm} {}+\nu_0 \, \Delta_{r_2}\Big[(r_2\!+\!1)N(r_1,r_2+1,t)
  + \nonumber\\ & & \hspace*{-0.5cm} 
  {}-({\cal A}-1)(L-r_2) N(r_1, r_2,t)\Big] \,.
\end{eqnarray}
In the continuum limit, Eq.~(\ref{2Ddiscrete_evolution}) becomes a 
two-dimensional generalization of the (biased) diffusion equation 
(\ref{general_evol}), 
\begin{eqnarray}
  \label{2DcontinuumEq}
  \frac{\partial}{\partial t}n(r_1,r_2,t) &=& \overline{\varphi}(r_1,r_2) 
  \,n(r_1,r_2,t) + \nonumber\\ & & \hspace*{-2.5cm} 
  {}+\frac{\partial}{\partial r_1}\left[D(r_1)\frac{\partial}{\partial r_1} 
  n(r_1,r_2,t) - v(r_1) n(r_1,r_2,t)\right]+\nonumber\\ & & \hspace*{-2.5cm} 
  {}+\frac{\partial}{\partial r_2}\left[D(r_2)\frac{\partial}{\partial r_2} 
  n(r_1,r_2,t) - v(r_2) n(r_1,r_2,t)\right]\;, 
\end{eqnarray}
where $D(r)$ and $v(r)$ are still given by Eqs.~(\ref{D}) and (\ref{v}) 
and $n(r_1,r_2,t)$, $\overline{\varphi}(r_1,r_2)$ are the continuum 
generalizations of $N(r_1,r_2,t)$ and $\bph(r_1,r_2)$, respectively. 
Fig.~\ref{fig7} illustrates the two-dimensional (biased) diffusion 
dynamics that emerges from Eq.~(\ref{2DcontinuumEq}).
The fitness function has a high plateau or `mesa' at $r_1,\,r_2<r_0$, 
and two strips of lower fitness along the $r_1$ and $r_2$ axis. 
Hence selection tries to keep $r_1,\,r_2<r_0$.
Mutation, on the other hand, drives the distribution towards the average 
number of mismatches in a random binding sequence, $r_1=r_2=(\cA-1)L/\cA$, 
as indicated by the arrows in Fig.~\ref{fig7}.
We are interested in the stationary distribution $N_0(r_1,r_2)$ 
that arises as a balance between selection and mutation. 
Below we characterize the dependence of $N_0(r_1,r_2)$ on the effective 
selection pressure $\talpha=2\alpha/\nu$ and the synergism factor $\omega$ 
numerically by iterating Eq.~(\ref{2Ddiscrete_evolution}). 
However, we first anticipate the qualitative behavior of $N_0(r_1,r_2)$ 
using the understanding of the single site problem gained in the last 
section. 

\begin{figure}
\includegraphics[width=6cm]{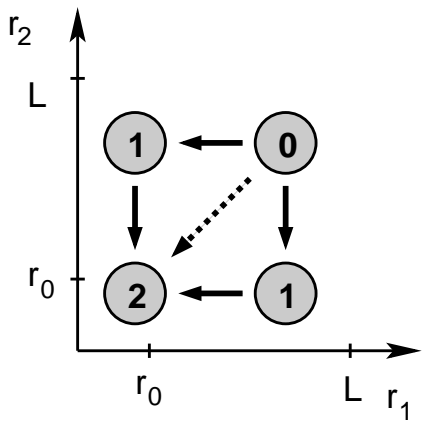}
\caption{\label{fig8}
Qualitative behavior of the stationary two-site mismatch distribution  
when the effective selection pressure $\talpha$ is increased. 
The circles indicate the position of the peak of the 
distribution, while the number inside the circle signifies the 
number of conserved motifs in the respective state. 
The dashed arrow indicates the direct pathway from the 0-motif state to 
the 2-motif state, while the solid arrows indicate the pathway 
through the intermediate 1-motif state (see discussion in the text). 
}
\end{figure}

Let us neglect a possible asymmetry between the sites for the moment, 
i.e. we set $\sigma=1$. 
It is clear that if $\talpha$ is below a certain threshold value, 
no motif will be maintained, i.e. the peak of the stationary 
distribution will be close to $r_1=r_2=(\cA-1)L/\cA$. 
On the other extreme, when $\talpha$ is very large, the distribution 
will certainly be localized on the high fitness mesa, corresponding to 
two conserved binding motifs. 
By analogy with the single site case, we would expect the distribution 
to be maximally fuzzy in this regime, and hence the peak of the 
stationary distribution to be close to $r_1=r_2=r_0$. 
What happens when $\talpha$ takes on intermediate values?
Upon increasing $\talpha$, the peak of the stationary distribution may 
either pass directly from $r_1=r_2=(\cA-1)L/\cA$ to $r_1=r_2=r_0$ or 
go through a state with only one conserved motif 
(see Fig.~\ref{fig8}). 
Intuitively, which of these ``pathways'' is taken, should depend on the 
value of $\omega$: when $\omega$ is small, the selective advantage of 
two conserved motifs over one is small and therefore a much higher 
selection pressure is needed to stabilize two motifs against mutations 
than just one, i.e. upon increasing the selection pressure the system 
passes from 0 to 1 to 2 motifs. 
Conversely, when $\omega$ is very large, two motifs are actually 
stabilized at lower selection pressures than a single motif would be, 
and hence the system passes directly from 0 to 2 motifs. 
We can estimate the value $\omega_c$ at which the system switches 
between the two pathways: 
when $(\omega-1)\ll 1$, the 1-motif phase exists in an intermediate 
range of $\alpha$'s, i.e. 
$\talpha_{c1}<\talpha<\talpha_{c2}$, where the lower critical value for 
the transition from 0 to 1 motif is approximately the same as in the 
single-site problem, i.e. $\talpha_{c1}\sim 1$, and the upper 
critical value is $\talpha_{c2}\sim (\omega -1)^{-1}$, since the 
transition from 1 to 2 motifs may be regarded as another single-site 
problem with $\alpha$ replaced by $(\omega-1)\alpha$. 
The system switches between the two pathways when 
$\talpha_{c1}=\talpha_{c2}$, and hence $\omega_c\approx 2$.

In the 1-motif phase, the selection of either motif, at site 1 or 
site 2, is equiprobable as long as $\sigma=1$. 
Correspondingly, $N_0(r_1,r_2)$ has two equal peaks as indicated in 
Fig.~\ref{fig8}. 
When $\sigma<1$ the peak associated with site 2 will disappear, but 
otherwise the qualitative behavior of the model remains the same. 
For simplicity, we will keep $\sigma=1$ from here on. 

\begin{figure}
\includegraphics[width=4cm]{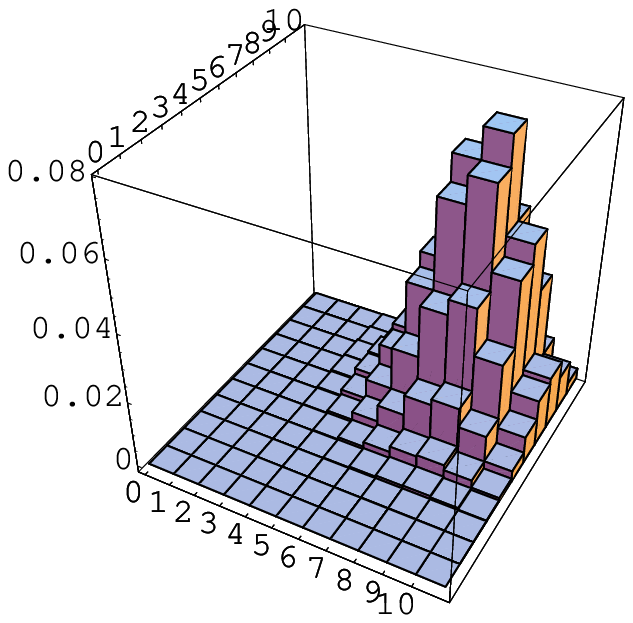}
\includegraphics[width=4cm]{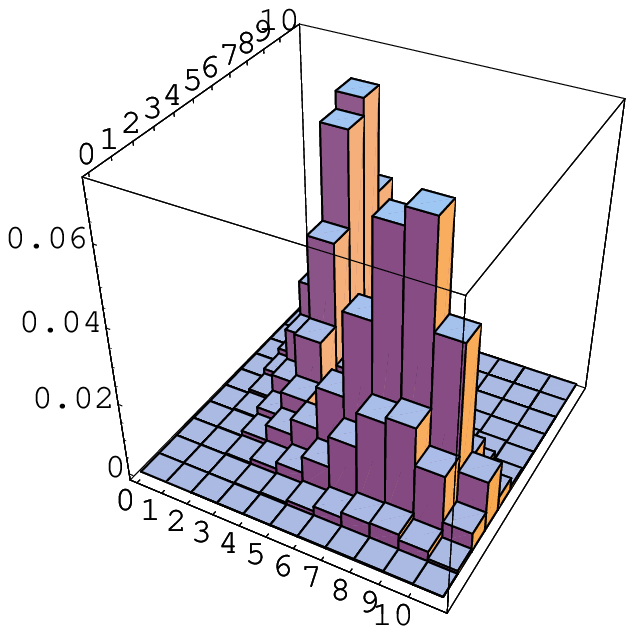}\\
\includegraphics[width=4cm]{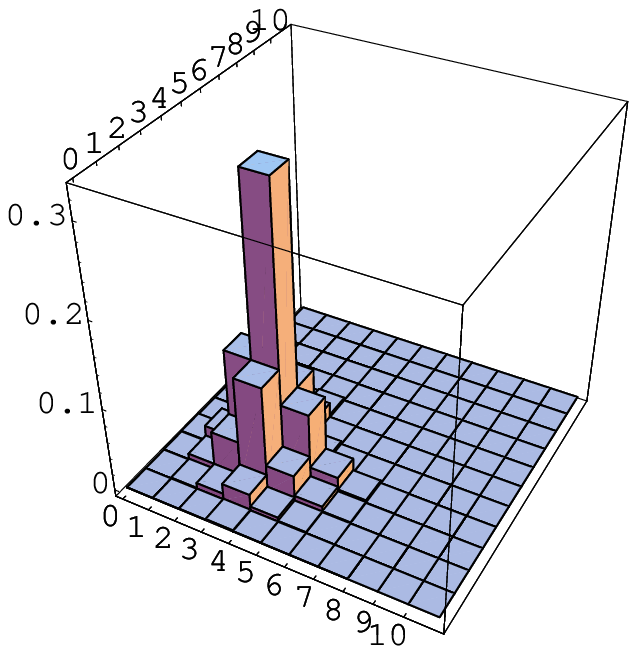}
\caption{\label{fig9}
Stationary distribution $N_0(r_1,r_2)$ in the 0-motif, 1-motif, 
and 2-motif phase ($\talpha=0.2$, 5, and 50, respectively, 
and $\omega=1.1$).}
\end{figure}

In order to examine the qualitative picture outlined above and to render 
it more quantitative, we now characterize the behavior of $N_0(r_1,r_2)$ 
numerically using the parameters tailored to CRP, i.e. $L=10$ and $r_0=3$. 
To determine $N_0(r_1,r_2)$, we evolve an arbitrary initial distribution 
$N(r_1,r_2,t=0)$ using Eq.~(\ref{2Ddiscrete_evolution}) until the 
stationary state is reached.
Fig.~\ref{fig9} displays three such stationary distributions, 
one each in the 0-motif, 1-motif, and 2-motif phase 
(here, we used $\talpha=0.2$, 5, and 50, together with $\omega=1.1$). 
Besides justifying the schematic sketch in Fig.~\ref{fig8}, 
it shows that the distributions both in the 1- and 2-motif phase are  
peaked at the ``edge'' $r_0=3$ and are therefore maximally fuzzy as in 
the single site problem. 

\begin{figure}
\includegraphics[width=8cm]{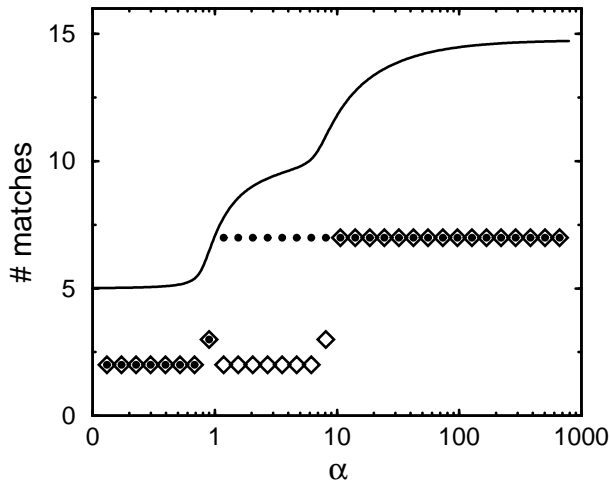}
\caption{\label{fig10}
Total number of matches in the two sites as a function of $\talpha$ 
(solid line), together with the number of matches in each site at 
the peak of the stationary distribution (first motif: dots, 
second motif: diamonds). 
[Note that in this figure the y-axis is reversed compared 
to Fig.~\ref{fig4}.]}
\end{figure}

Next, we focus on the transitions between the three phases. 
In Fig.~\ref{fig10}, the average total number of {\em matches}, 
i.e. $2L\!-\!\langle r_1\rangle\!-\!\langle r_2\rangle$ 
(here $\langle\ldots\rangle$ denotes an average over the stationary 
distribution), is plotted against $\talpha$, again with $\omega=1.1$ 
(solid line). [Note that in this figure the y-axis is reversed compared 
to Fig.~\ref{fig4}.]
We observe that the total number of matches rises quite sharply around 
$\talpha=1$ and $\talpha=10$. 
These positions agree with our estimates $\talpha_{c1}\sim 1$ 
and $\talpha_{c2}\sim (\omega -1)^{-1}$ based on the qualitative 
discussion above. 
(Note that since we work with a small `system size' of $L=10$, the 
transitions, which are sharp in the limit $L\to\infty$, appear only as 
smooth crossovers.) 
In order to show that the rises are indeed caused by the transitions 
from the 0-motif to the 1-motif phase, and from the 1-motif to the 
2-motif phase, respectively, we also plotted the number of matches 
in each site at the peak (at one of the two peaks in the 1-motif 
phase) of the stationary distribution in Fig.~\ref{fig10} 
(circles and triangles). 
This illustrates the typical behavior of the individual sites: 
the first site jumps from $2\sim 3$ matches to 7 matches at 
$\talpha\approx 1$ and the second site does the same at 
$\talpha\approx 10$. 
Hence, the motifs clearly appear in a step-like fashion as the 
selection pressure is increased. 

In evolution experiments with RNA viruses, this twofold transition 
should be directly observable (if a suitable operon can be identified 
where the fitness function (\ref{2site_fitness}) is applicable), 
since according to our estimates above, 
$\talpha$ for these systems can be tuned over a range of $1\sim 100$ 
by varying the fractional exposure to different environmental conditions. 
On the basis of our model, one would expect, for instance, that one of the 
sites in a doublet disappears in the course of an evolution experiment, 
if the selection pressure is sufficiently lowered by reducing the 
exposure to the environment where binding is beneficial. 
When the exposure is reduced to zero, both regulatory sites 
{\em and} the gene coding for the transcription factor (if not required 
for other mechanisms) will be lost. 

\begin{figure}
\includegraphics[width=8cm]{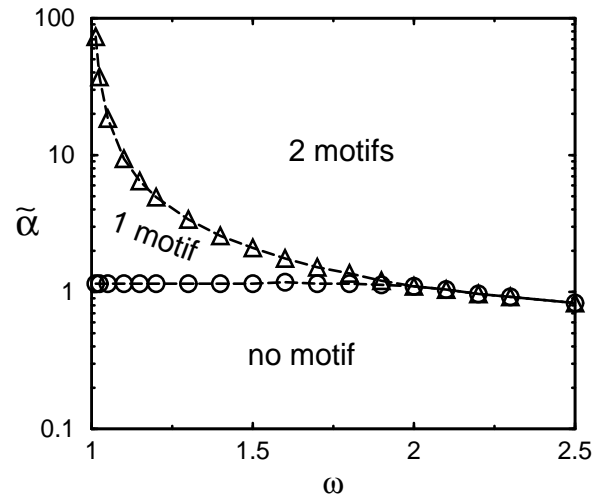}
\caption{
Phase diagram in $(\alpha,\omega)$-space. 
The boundaries are given by $\talpha_{c1}$ (circles) and 
$\talpha_{c2}$ (triangles) as a function of $\omega$.}
\label{fig11}
\end{figure}

To complete our characterization of the model behavior, we map out the 
entire phase diagram in the $(\talpha,\omega)$ parameter-space. 
The result is shown in Fig.~\ref{fig11}, where 
$\talpha_{c1}$ and $\talpha_{c2}$ are plotted as a function of $\omega$.
Since $L=10$ in the present case and the phase boundary is well-defined 
only in the limit $L\to\infty$, the curves $\talpha_{c1}(\omega)$ and 
$\talpha_{c2}(\omega)$ are really only crossover lines whose precise 
location is slightly dependent on their definition (in the figure they 
are represented by dashed lines in order to indicate this fact). 
[Here, we defined $\talpha_{c1}$ and $\talpha_{c2}$ as the value of 
$\talpha$ where the peak of the stationary distribution first reaches 
7 matches in the respective site.] 
We see that the phase boundaries join at $\omega\approx2$ as expected 
from our estimate above. 
When $\omega\to 1$ from above, the upper boundary, $\talpha_{c2}$, 
diverges as $(\omega-1)^{-1}$. 
This implies that at $\omega=1$, the two-motif phase is unreachable, 
regardless of the value of $\talpha$. 
Finally, we note that \citet{hermisson} observe a somewhat similar 
two-fold transition between three phases in a situation where two 
different mutation processes with distinct mutation rates are taken 
into account. 

Let us now return to the scenario outlined at the beginning of this 
section, and discuss whether we can interpret the regulatory regions 
with multiplets as being under higher selective pressure for factor 
binding than the ones with single binding sites. 
Our study of the two-site problem would justify this conclusion, 
if (a) the values of $\omega$ and $\sigma$ 
were very similar for all regulatory regions, 
and (b) the effective selection pressure were typically on the same 
order of magnitude as $\talpha_{c2}$. 
Then, we could tentatively associate singlets with an $\talpha$ below 
$\talpha_{c2}$, and multiplets with an $\talpha$ above that threshold. 
However, both conditions are not likely to be fulfilled generically 
in bacteria. 
First, the values of $\omega$ and $\sigma$ not only depend on the sequence 
of the promoter and the distance of the binding sites from the 
promoter, but also on the level of gene expression that is beneficial 
for the cell. 
For example, genes that code for proteins which are not needed in 
large amounts, even in the environmental condition where their 
expression is favored, do not require a large activation, and hence 
the effect of a second binding site could even be detrimental, i.e. 
$\omega<1$. 
Second, $\talpha$ should typically be on the order of $10^4$ 
(see above), and hence, as long as $\omega$ is only slightly 
larger than one ($\omega>1.0001$), our model would always predict 
multiple binding sites for bacterial transcription factors. 
Therefore, within our model, 
whether one or two motifs are maintained depends almost 
exclusively on the value of $\omega$, i.e. $\omega\le1$ leads to one 
motif and $\omega>1$ to two motifs. 
However, there may be cases where maintaining subtle differences in  
the temporal ordering of turning on/off different operons 
would give rise to a very small fitness advantage, e.g. 
flagella assembly and SOS-respnse in {\it E. coli} 
(see recent results by U. Alon, submitted for publication).
In such cases, the system may respond by keeping one or two motifs 
according to the theory we presented.
And of course there is also the situation of RNA viruses described 
above where the twofold transition as depicted in 
Fig.~\ref{fig10} could in principle be directly 
observed. 
\section{Discussion and Conclusions}
The fuzziness of regulatory binding motifs is a widely observed 
phenomenon. 
The present investigation has shown that the entropic advantage 
of introducing mismatches from the best binding sequence is 
sufficient to produce motifs that are maximally fuzzy while still 
retaining the capability of factor binding. 
Nevertheless, we cannot exclude that the fuzziness actually bears 
a selective advantage (in the language of population genetics, 
this would correspond to a stabilizing selective pressure).  
The alternative scenario given for the fuzziness of the CRP sites
is an explicit example of the latter.
It would be very interesting to address this question experimentally 
by directly measuring the fitness of a bacterium or virus as a 
function of the sequence of its binding sites: 
starting with a wild-type binding site that has several mismatches, 
what is the effect on the fitness, when the number of 
mismatches is reduced by site-directed mutagenesis? 
Does the fitness remain unchanged or is it reduced?
Besides answering the question raised above, experiments of that 
type could also lead to important conclusions on the evolution of 
genetic regulation. 

Another important result of our study is the phase transition 
associated with the maintenance of motifs. 
Our general conclusion is that the selection pressure on a single 
binding motif needs to surpass a threshold value of approximately 
$\nu_0 L/2$ to guarantee maintenance, while the threshold for a 
second site (for the same factor and in the same regulatory region) 
is larger by a factor $(\omega-1)^{-1}$, where $\omega$ is given 
by the ratio of the fitness of the organism with two sites over the 
fitness of the organism with one site. 
As pointed out above, this prediction could be tested experimentally 
by evolving RNA viruses in a fluctuating environment and varying 
the fractional exposure to the environment where factor binding is 
beneficial. In this case there would be no need to do site-directed
mutagenesis, since the transition could be observed directly by 
sequencing. 

Our model makes quantitative predictions on the 
{\it statistics of polymorphisms} in binding sites. 
To test these, it will not suffice to sequence a particular binding 
site in many different isolates from a single, large ($N\nu_0\gg 1$) 
laboratory population, since this population originates from a small, 
genetically homogeneous population and it takes a time on the order of 
$\nu_0^{-1}$ to equilibrate the distribution of mismatches in a binding 
site. Instead, sequencing the same binding sites in several different 
strains should yield the desired data. 
Besides allowing a comparison to our model, detailed information on 
polymorphisms in binding sequences would also make it 
possible to address a number of interesting questions, e.g.  
how does the selection pressure on binding sites compare with the selection 
pressure on coding sequences\footnote{For example, we would expect that the 
selection pressure on the coding sequence of the binding region in 
transcription factors such as 
crp or lexA, which have many binding sites distributed over the whole 
genome, is much higher than on individual operator sites, since a 
change in the sequence of an operator site affects only the regulation of 
that particular site, while a change in the aminoacid sequence of the binding 
region of the protein affects the regulation of many genes.}? 
Or, can one identify compensating mutations between promoter and binding site 
sequences, e.g. could a mutation that weakens the promoter be compensated by 
a mutation that increases the affinity of an activator to its operator site? 

We conclude that the evolution of transcription factor binding sites is a 
problem that is well suited to establishing a link between the detailed 
molecular biophysics of a system and its evolutionary dynamics. 
The theory presented in the present article is meant as a first step,
with the hope of stimulating future experiments in this direction.
There are many directions for the improvement of the model and the analysis.
One important question is the validity of the mean-field analysis
described here. Is the finite population size effect important here and
how would it change the motif statistics within our model?
One can also investigate more elaborate models including, for instance, 
the effect of time-dependent environment, 
the coupled evolution of the polymerase and transcription factor
the binding sites, and the interaction among different factors.
\section*{Acknowledgements}
It is a pleasure to acknowledge useful discussions with 
Carson Chow, David Moroz, Luca Peliti and David Thaler. This work
is supported in part by the National Science Foundation through
Grant Nos. DMR-9971456 and DBI-9970199. UG is supported in part by
a German DAAD fellowship, and TH by a Burroughs Wellcome functional
genomics award.

{\bf Note added:} Upon completion of the present work,
we learned of the work of \citet{sengupta} who
also examined the protein-DNA interaction from an evolutionary 
perspective. While very similar mean-field evolution equations
are analyzed in both works, the emphasis are quite different,
with \citet{sengupta} arguing for a ``robustness'' criterion 
based on minimizing mutational load.
\end{document}